\documentclass[12pt]{article}
\usepackage{graphicx,amssymb,amsmath,cite}
\usepackage[dvips]{epsfig}
\input epsf.tex

 \newlength{\wth}
\def\slashchar#1{\setbox0=\hbox{$#1$}
   \dimen0=\wd0
   \setbox1=\hbox{/} \dimen1=\wd1
   \ifdim\dimen0>\dimen1
      \rlap{\hbox to \dimen0{\hfil/\hfil}}
      #1
   \else
      \rlap{\hbox to \dimen1{\hfil$#1$\hfil}}
      /
   \fi}

\newcommand{\lsim}{\;\raisebox{-0.9ex}{$\textstyle\stackrel{\textstyle<}
           {\sim}$}\;}

\begin{document}
\begin{titlepage}

\begin{center}
{\tiny  {CAVENDISH-HEP-2009-09,} {DAMTP-2009-42,} {LPSC-09-114,} {SCUPHY-TH-09003} } \\
\end{center}

\bigskip\bigskip

\begin{center}{\Large\bf\boldmath
Neutralino Reconstruction at the LHC from Decay-frame Kinematics}
\end{center}
\bigskip
\centerline{\bf Z. Kang and N. Kersting} \centerline{{\it Physics
Department, Sichuan University }} \centerline{{\it Chengdu, Sichuan
Province, P.R. China 610065}}
\bigskip
\centerline{\bf S. Kraml} \centerline{{\it Laboratoire de Physique
Subatomique et de Cosmologie (LPSC) }} \centerline{{\it   UJF
Grenoble 1, CNRS/IN2P3, 53 Avenue des Martyrs, F-38026 Grenoble,
France}}
\bigskip
\centerline{\bf A.R. Raklev} \centerline{{\it DAMTP, Wilberforce
Road, Cambridge, CB3 0WA, UK }} \centerline{{\it Cavendish
Laboratory, JJ Thomson Avenue, Cambridge, CB3 0HE, UK}}
\bigskip
\centerline{\bf M.J. White} \centerline{{\it Cavendish Laboratory,
JJ Thomson Avenue, Cambridge, CB3 0HE, UK }}
\bigskip

\begin{abstract}
Decay-frame Kinematics (DK) has previously been introduced as a
technique to reconstruct neutralino masses from their three-body
decays to leptons. This work is an extension to the case of two-body
decays through on-shell sleptons, with Monte Carlo simulation of LHC
collisions demonstrating reconstruction of neutralino masses for the
SPS1a benchmark point.

\end{abstract}

\newpage
\pagestyle{empty}

\end{titlepage}

%%%%%%%%%%%%%%%%%%%%%%%%%%%%%%%%%%%%%%%%%%%%%%%%%%%%%%%%%%%%%%%%%%%%%%

\section{Introduction}

The Large Hadron Collider (LHC) at CERN, Geneva, is expected to
provide direct evidence for any New Physics beyond the Standard
Model (SM) at the TeV energy scale. The properties of these
particles may shed light on the origin of electroweak symmetry
breaking and the nature of dark matter, and give clues to a more
fundamental theory of Nature. To explain the deficiencies of the SM,
a large variety of theories has been put forward, and it is only by
carefully measuring the properties of any new particles that one
will be able to discriminate between them.  Chief amongst these
properties are the masses of the new particles.

Many of these new theories ({\it e.g.}, supersymmetry and extra dimensions)
contain a stable weakly interacting massive particle that will only be
`visible' in an LHC detector as missing energy. Such particles are
natural dark matter candidates by virtue of their interactions, but
pose problems for mass measurements at hadron colliders since one
cannot in general reconstruct the kinematics of an event in which
these particles are produced. To worsen the problem, the
parton--parton interactions in the collider have by their very nature
an unknown center of mass energy.

In the literature a number of techniques has been developed to get
around this problem. These fall into two general classes: those that
perform a fit to or set a limit on masses using information from the
entire event sample, and those which rely exclusively on events near
the endpoint of a kinematic distribution. Belonging to the first class
are ``Mass-Shell Techniques" (MSTs), represented by the work done in
\cite{Kawagoe:2004rz,Nojiri:2007pq},
\cite{Cheng:2007xv,Cheng:2008mg,Cheng:2009fw}
and \cite{Webber:2009vm},
which depend on maximizing the solvability of assumed
mass-shell constraints in a given sample of events. This has been
shown to be very effective if enough such constraints are
available.\footnote{However, problems arise if, {\it e.g.}, there are
three-body decays or too many invisible decay products;
see~\cite{Bisset:2008hm} for further discussion.} Also belonging to
this class are techniques which work with extrema of a ``transverse
mass" variable, {\it e.g.},
$m_{T2}$~\cite{Lester:1999tx,Barr:2003rg,Cho:2007qv,Barr:2007hy,
Cho:2007dh,Nojiri:2008hy,Barr:2008ba}, and techniques which look at
the shape of complete invariant mass
distributions~\cite{Miller:2005zp, Kraml:2005kb,Gjelsten:2006as,
Gjelsten:2006tg,Kraml:2008zr}. Much research has recently become
focused in these areas.

The second class of mass reconstruction techniques most notably
includes the traditional kinematic endpoint method
\cite{Hinchliffe:1996iu,Paige:1997xb,Bachacou:1999zb,Lytken,Gjelsten:2004ki,
Gjelsten:2005aw, Huang:2008qd,Burns:2009zi,Matchev:2009iw},
where the endpoints of various
invariant mass distributions can be matched to analytical functions of
the unknown masses, that in turn can be inverted to solve for the same
masses in suitably long decay chains.  Such methods have been studied
for over a decade already, and would seem to have been largely
explored for the simplest, and most probable, decay chains in popular
theories such as the Minimal Supersymmetric Standard Model
(MSSM). However, there is more to be done with events near an endpoint
as demonstrated recently in~\cite{Kersting:2009ne}: here a Decay-frame
Kinematics (DK) technique utilizes the fact that events at a kinematic
endpoint \emph{can} have exactly-known kinematics in terms of
production angles and energies of all particles in the assumed decay
chain. Events near an endpoint will thus have approximately known
decay-frame kinematics, which allows one to constrain and solve for
unknown masses.  In~\cite{Kersting:2009ne} this was demonstrated for
the case of neutralino three-body decays through off-shell sleptons to
lepton pairs plus missing energy carried away by the lightest
supersymmetric particle (LSP), the lightest neutralino, {\it e.g.\
}$\tilde\chi_2^0\to\ell^+\ell^-\tilde\chi_1^0$. The on-shell case
was deferred to a future work --- this work.

In the following we will demonstrate the use of DK in the case of
on-shell neutralino decays, {\it i.e.\ }$\tilde\chi_i^0
\to\tilde\ell^\pm\ell^\mp\to\ell^+\ell^-\tilde\chi_1^0$, though it
should be stressed that the technique demonstrated can be applied to
any similar cascade decay process arising in any New Physics model.
Section \ref{sec:off} begins with a review of the off-shell case,
demonstrating its application at a NMSSM parameter point. Section
\ref{sec:on} then discusses the main new development of this paper,
the generalisation of the DK technique to the on-shell case, where it
is found that several subtleties emerge beyond what was found for the
comparatively simple off-shell case; we further present a Monte Carlo
(MC) study of the mSUGRA SPS1a benchmark point~\cite{Allanach:2002nj},
where the DK technique proves quite capable of reconstructing the
relevant neutralino masses from $\tilde\chi_2^0\tilde\chi_2^0$ decays.
Section \ref{sec:conc} gives our conclusions.

\section{Off-Shell Decays}
\label{sec:off}

\begin{figure}[!t]
\begin{center}
\includegraphics[width=0.49\linewidth]{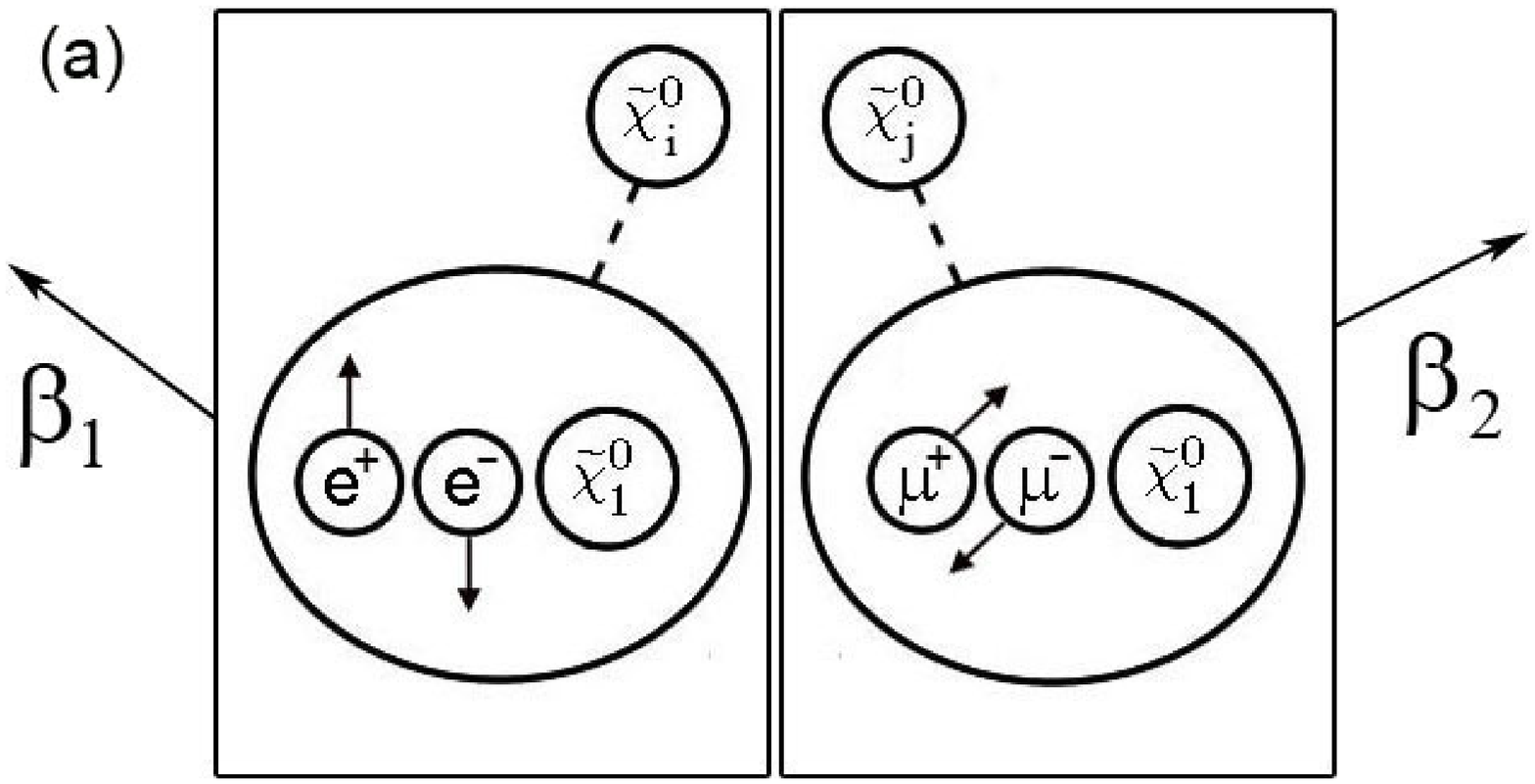}
\includegraphics[width=0.49\linewidth]{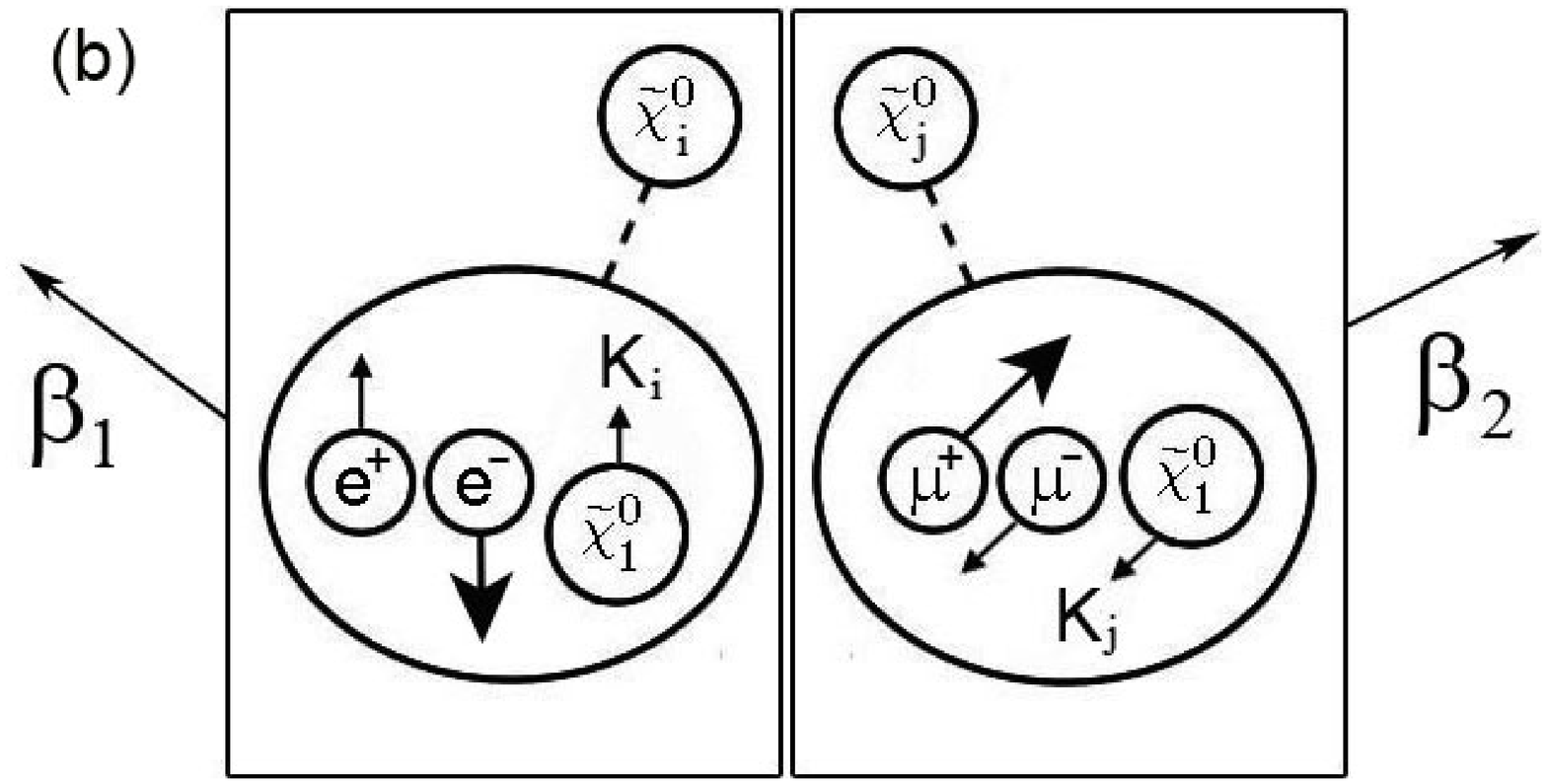}
\end{center}
\caption{\small \emph{(a) Neutralino three-body decays with maximal $M_{ee}$ and
$M_{\mu\mu}$: though the decaying $\tilde\chi_{i,j}^0$ may be moving
with any velocity $\beta_{1,2}$ in the lab frame, in each respective
decay frame the leptons have equal and opposite momenta while the
$\tilde\chi_1^0$ (LSP) is at rest. (b) If the $\tilde\chi_{i,j}^0$ has
a two-body decay via on-shell sleptons, the LSPs are no longer
stationary in the neutralino decay frame, but have momenta $K_{i,j}$
collinear with the leptons' momenta.}}
\label{fig:zizj}
\end{figure}

We begin with a brief review of the DK technique applied to the case
of neutralino three-body decays. For a more complete treatment, see
\cite{Kersting:2009ne}. We consider production of neutralino pairs in
the MSSM which undergo three-body decays to electrons, muons, and
$\tilde\chi_1^0$ (the LSP):
\begin{equation}
\label{zizjdecay2}
pp \to \mathbb{X}\to \mathbb{X}' + \tilde\chi_i^0
(\to e^+ e^- \tilde\chi_1^0)~\tilde{\chi}_{j}^0 (\to\mu^+\mu^-\tilde\chi_1^0),
\end{equation}
where $\mathbb{X}$ represents either a $Z^*$ or any MSSM production
channel via a Higgs ($H^0$ or $A^0$) or cascade from gluino/squark
pair-production, while $\mathbb{X}'$ are SM states potentially
produced in association, relevant in this context only for the
measurement of missing momentum.  The physical observables of interest
from one event thus consist of four leptonic four-momenta
$p_{e^\pm,\mu^\pm}$, from which we may construct the usual di-lepton
invariant masses, $M_{ee}$ and $M_{\mu\mu}$, and missing momentum in
two transverse directions, assumed equal to the sum of the two LSPs'
transverse momenta, $p_{\chi,\chi'}^T$. If we happen to have an event
where both the invariant masses $M_{ee}$ and $M_{\mu\mu}$ are maximal,
as shown in Fig.~\ref{fig:zizj}a, it will be subject to the system of
constraints (hereafter we abbreviate $m_i
\equiv m_{\widetilde{\chi}_{i}^0}$)
\begin{eqnarray}
\label{con1}
M_{ee}^{\max}     & = & m_i - m_1, \\ \label{con2}
M_{\mu\mu}^{\max} & = & m_j - m_1, \\ \label{con345}
\vec{p}_{e^+}^{~\prime} + \vec{p}_{e^-}^{~\prime} & = & 0, \\ \label{con678}
\vec{p}_{\mu^+}^{~\prime} + \vec{p}_{\mu^-}^{~\prime} & = & 0, \\ \label{con910}
(\vec{p}_\chi + \vec{p}_{\chi'})^T & = & \not\!{\vec{p}}^{~T}~\mathrm{(observed),}
\end{eqnarray}
where leptonic momenta are written in the rest frame of the respective
parent neutralino, and $\vec{p}_{e^\pm}^{~\prime}=\mathbf{\Lambda}_1
\vec{p}_{e^\pm}$ and $\vec{p}_{\mu^\pm}^{~\prime}=\mathbf{\Lambda}_2
\vec{p}_{\mu^\pm}$ define the appropriate Lorentz transformations
$\mathbf{\Lambda}_{1,2}$ from the lab frame. This system gives ten
equations for the nine unknowns, the velocities $\vec\beta_{1,2}$ and
the masses $m_{1,i,j}$, allowing us to actually overconstrain the
masses $m_{1,i,j}$. The $\vec{\beta}_{1,2}$ which satisfy
(\ref{con345}) and (\ref{con678}), making the total momentum of each
lepton pair zero, are uniquely given by
\begin{equation}
\label{betaeqn1}
\vec\beta_1 = \frac{\vec{p}_{e^+} + \vec{p}_{e^-}}{E_{e^+} + E_{e^-}}
\quad{\rm and}\quad
\vec\beta_2 = \frac{\vec{p}_{\mu^+} + \vec{p}_{\mu^-}}{E_{\mu^+} + E_{\mu^-}}.
\end{equation}

Now, because of the condition that $M_{ee}$ and $M_{\mu\mu}$ are
maximal, the corresponding $\mathbf{\Lambda}_{1,2}$, which take the
$e^+ e^-\tilde\chi_1^0$ and $\mu^+ \mu^- \tilde\chi_1^0$ systems to
their respective $\tilde\chi_{i,j}^0$ rest frames, also bring each
$\tilde\chi_1^0$ to rest. Thus their four-momenta in these frames must
be $(m_1,\vec{0})$, which, when inverse-Lorentz-transformed by
$\mathbf{\Lambda}_{1,2}^{-1}$, giving $(m_1\gamma_{1,2}~,m_1
(\vec\beta\gamma)_{1,2} )$, have to agree with the observed missing
momentum $\not\!\vec{p}^{~T}$; this matching condition along each
transverse direction (say $\hat{x}$ and $\hat{y}$) then gives two
independent determinations of $m_1$:
\begin{equation}
\label{m1eqn}
m_1' = \frac{\not\!{p}_x}{(\beta_x\gamma)_1 +
  (\beta_x \gamma)_2}
\quad{\rm and}\quad
m_1'' = \frac{\not\!{p}_y}{(\beta_y\gamma)_1 + (\beta_y\gamma)_2}.
\end{equation}

Since we are assuming that both $M_{ee}$ and $M_{\mu\mu}$ are
precisely maximal --- the perfect event of Fig.~\ref{fig:zizj}a ---
we should get $m_1' = m_1''= m_1$ from such an event. In practice of
course, we can only expect to find an event within some neighborhood
$\epsilon$ of the endpoints, $M_{ee,\mu\mu}=M_{ee,\mu\mu}^{\max}\pm
\epsilon$, in which case one finds that $m_1'$ and $m_1''$ are
offset by ${\mathcal O}(\sqrt{2\epsilon m_1})$ from $m_1$
\cite{Kersting:2009ne}. One might then expect that applying
(\ref{betaeqn1}) and (\ref{m1eqn}) to a sample of events near the
endpoint should give a distribution of $m_1'$ and $m_1''$ peaked
near $m_1$ with a spread determined by sample purity.

Here, to lend further support to the generality of the above
technique, let us demonstrate its application to the rather
challenging example of an NMSSM (Next-to-Minimal Supersymmetric
Standard Model) scenario described in~\cite{Kraml:2008zr}. This has
a supersymmetric particle spectrum containing five neutralinos, the
lightest of which is 99\% singlino, and a generic feature of the
parameter space that gives the correct dark matter density is a
significant degeneracy between the singlino and second lightest
neutralino mass. This gives rise to copious production of soft
lepton pairs with small invariant masses, $M_{\ell\ell} \lsim
10$~GeV, from the three-body decay
$\tilde\chi_2^0\to\ell^+\ell^-\tilde\chi_1^0$. For more details on
this scenario see~\cite{Kraml:2008zr}.

\begin{figure}[h]
\begin{center}
\includegraphics[width=0.32\linewidth]{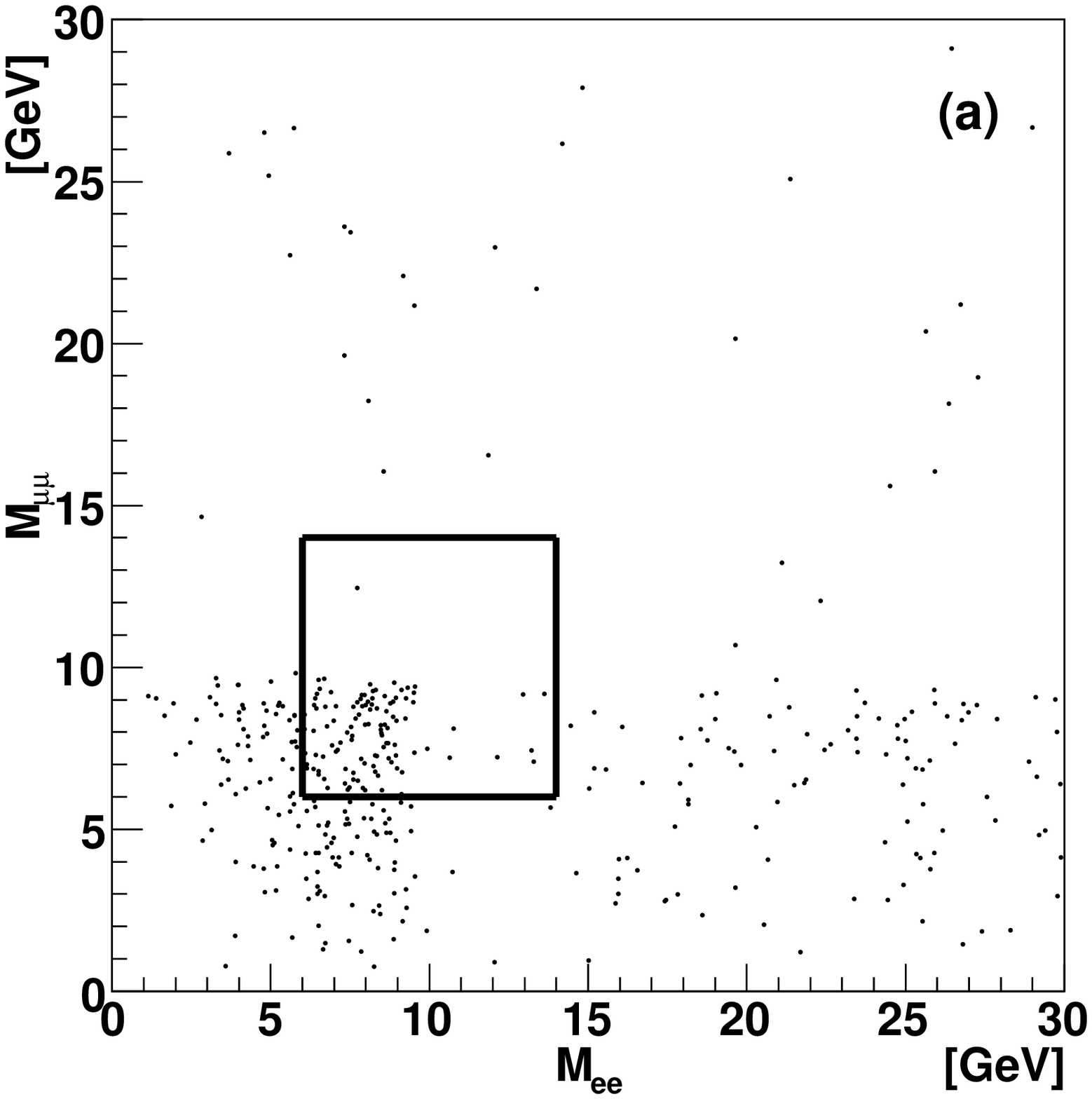}
\includegraphics[width=0.32\linewidth]{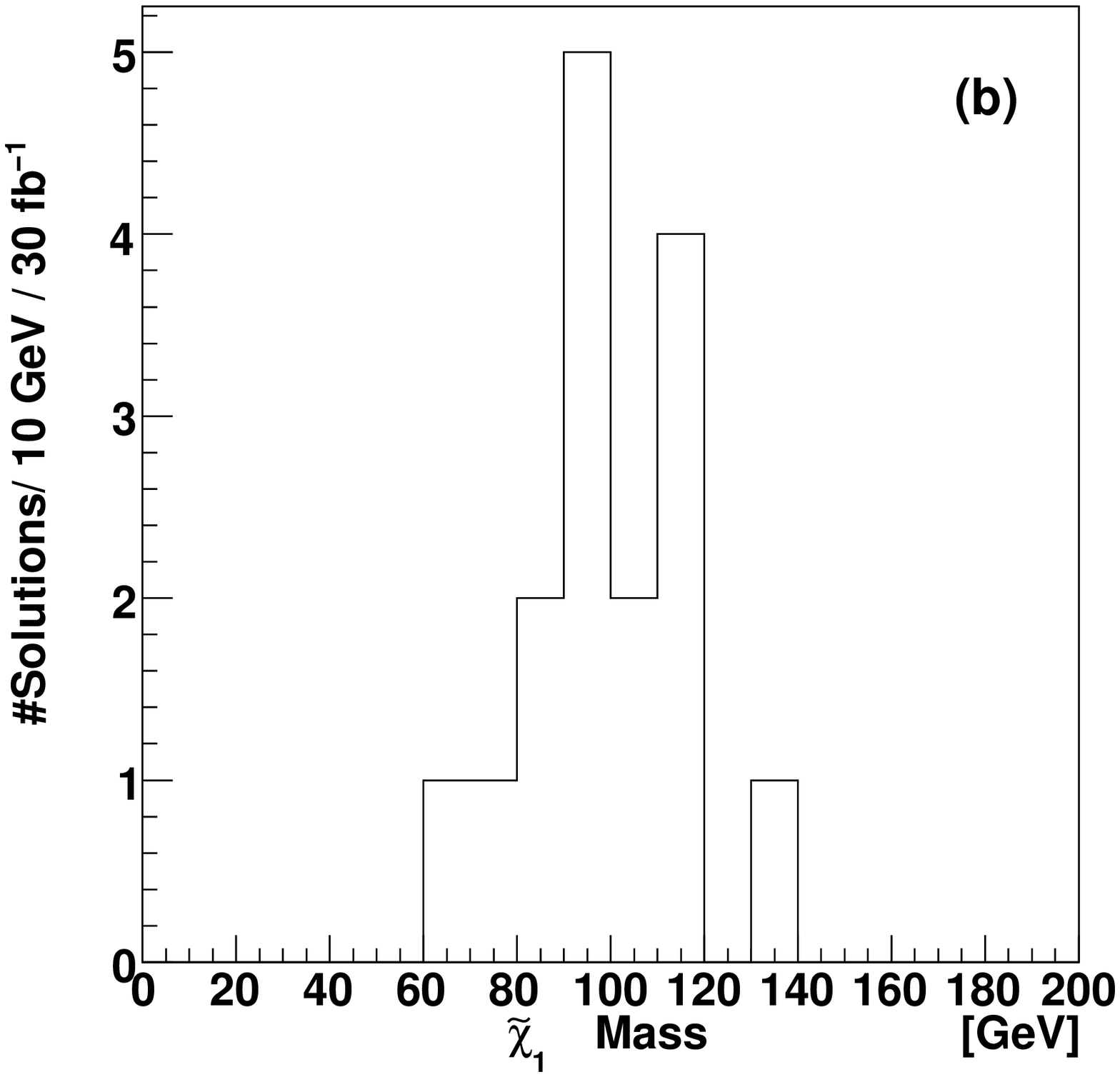}
\includegraphics[width=0.32\linewidth]{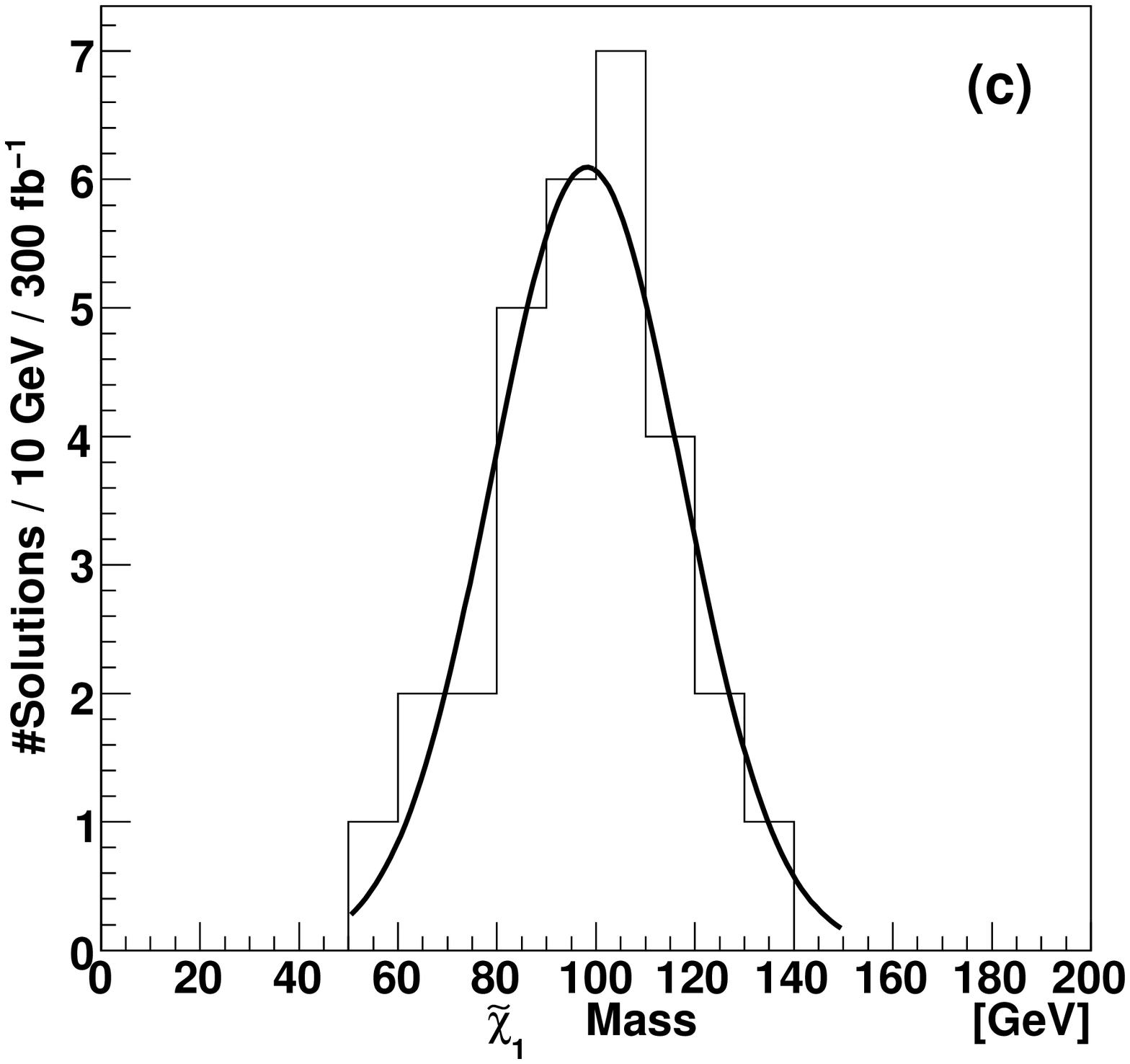}
\end{center}
\caption{\small \emph{ (a) Wedgebox plot for 30~fb$^{-1}$ of integrated
LHC luminosity for ``Point A" of \cite{Kraml:2008zr}. (b) The
$\tilde\chi_1^0$ mass distribution for events sampled from the boxed
region in (a). We see a peak in rough agreement with the nominal value of
$m_1 = 105.4$~GeV. (c) Same as (b) but for 300~fb$^{-1}$ and
$\epsilon=1$~GeV.
}}
\label{fig:offdk}
\end{figure}

We study the benchmark ``Point A" of that paper, a point which has
$M_{\ell\ell}^{\max}=9.7$~GeV and $m_1=105.4$~GeV, using the same MC
setup and fast detector simulation as in~\cite{Kraml:2008zr}. For
details of the simulation see also Section~\ref{sec:on} of the present
paper. To isolate signal events of the type (\ref{zizjdecay2}), we
place the following cuts on our events:
\begin{itemize}
\item
Require missing transverse energy $\not\!\!E_T>100$~GeV.
\item Require at least two hard jets with $p_T>150,\,100$~GeV.
\item
Require four isolated leptons with flavor structure $e^+e^-\mu^+\mu^-$
and $p_T>7\,(4)$ for $e$ $(\mu)$. All such leptons must pass the lepton
efficiency cuts employed in \cite{Kraml:2008zr} modeled on full
simulation results given in \cite{:2008zzm}.
\end{itemize}
From the surviving events we construct a wedgebox plot of the
di-electron versus the di-muon invariant mass, for a number of events
equivalent to 30~fb$^{-1}$ of statistics at the LHC. The result is
seen in Fig.~\ref{fig:offdk}a, showing a clear box-like structure at
$M_{ee,\mu\mu}\sim 10$~GeV, the endpoint of the di-lepton invariant
mass distribution for the $\tilde\chi_2^0$ decay.

Choosing a sampling region in a rather generous neighborhood of the
endpoint, $M_{\ell\ell} = 10\pm 4$~GeV gives ${\mathcal O}(100)$
events. We now apply Eqs.~(\ref{betaeqn1}) and (\ref{m1eqn}) to each
of these, demanding that $m_1'$ and $m_1''$ agree to within 20\%.
Although only about 20 events survive this criterion, the resulting
$m_1$ distribution can be seen in Fig.~\ref{fig:offdk}b to peak
quite prominently slightly below the nominal value of $m_1 =
105.4$~GeV. The systematic error of the method is seen to be
comparable to the estimate made earlier.

With higher statistics, this allows us to determine the absolute LSP
mass to rather good precision. The results for 300~fb$^{-1}$ of data
are shown in Fig.~\ref{fig:offdk}c. Here we narrow down our sampling
region to $\epsilon =1$~GeV from the wedgebox edge. With a Gaussian
distribution we obtain a best fit value of $m_1=98.2\pm 3.9$~GeV
with $\chi^2/{\rm ndf}=0.41$.

\section{On-Shell Decays}
\label{sec:on}

\subsection{Kinematics}
\label{subsec:kin}

Let us now continue to the main focus of this paper, {\it i.e.\ }the
added complications that arise when the neutralinos decay through
on-shell intermediate sleptons:
\begin{equation}
\label{hdecay}
\mathbb{X} \to \mathbb{X}' + \tilde\chi_i^0 \tilde\chi_j^0
(\to e^\pm\tilde e^\mp\mu^\pm\tilde\mu^\mp
\to e^+ e^- \mu^+ \mu^- \tilde\chi_1^0\tilde\chi_1^0).
\end{equation}
When $M_{ee}$ and $M_{\mu\mu}$ are maximal, as illustrated in
Fig.~\ref{fig:zizj}b, two-body kinematics gives the following system
of constraints:
\begin{eqnarray}
\label{con1b}
M_{ee}^{\max}     & = & m_i \sqrt{1 - (m_s/m_i)^2} \sqrt{1 - (m_1/m_s)^2}, \\ \label{con2b}
M_{\mu\mu}^{\max} & = & m_j \sqrt{1 - (m_s/m_j)^2} \sqrt{1 - (m_1/m_s)^2}, \\ \label{con34b}
\vec{p}_{e^+}^{~\prime}   & \parallel & -\vec{p}_{e^-}^{~\prime},   \\ \label{con56b}
\vec{p}_{\mu^+}^{~\prime} & \parallel & -\vec{p}_{\mu^-}^{~\prime}, \\ \label{con7b}
|\vec{p}_{e^+}^{~\prime}+\vec{p}_{e^-}^{~\prime}| & = &
\left|\frac{m_s^4 - m_i^2m_1^2}{2m_im_s^2} \right| \equiv K_i, \\ \label{con8b}
|\vec{p}_{\mu^+}^{~\prime}+\vec{p}_{\mu^-}^{~\prime}| & = &
\left|\frac{m_s^4 - m_j^2m_1^2}{2m_jm_s^2}\right| \equiv K_j, \\  \label{con910b}
(\vec{p}_\chi + \vec{p}_{\chi'})^T & = & \not\!\vec{p}^{~T}~\mathrm{(observed)},
\end{eqnarray}
where we have assumed a common slepton mass $m_s=m_{\tilde{e}} =
m_{\tilde\mu}$. See the Appendix for details of the derivation. The
antiparallel conditions (\ref{con34b}) and (\ref{con56b}) force
$\vec{\beta}_{1,2}$ to be in the planes of the respective leptons, so
there are really only four boost parameters to find; adding the
unknown masses $m_{1,i,j,s}$ to this gives eight unknowns which can
thus be solved for by the eight constraints
(\ref{con1b})--(\ref{con910b}).  In principle, if one were handed an
event of the type in Fig.~\ref{fig:zizj}b, one could numerically apply
(\ref{con1b})--(\ref{con910b}), scanning over the eight-dimensional
space of unknowns for a solution. Needless to say, this is not the
most practical approach, nor particularly enlightening as to the nature
of any solution which might be found.

\begin{figure}[!htb]
\begin{center}
\includegraphics[width=0.49\linewidth]{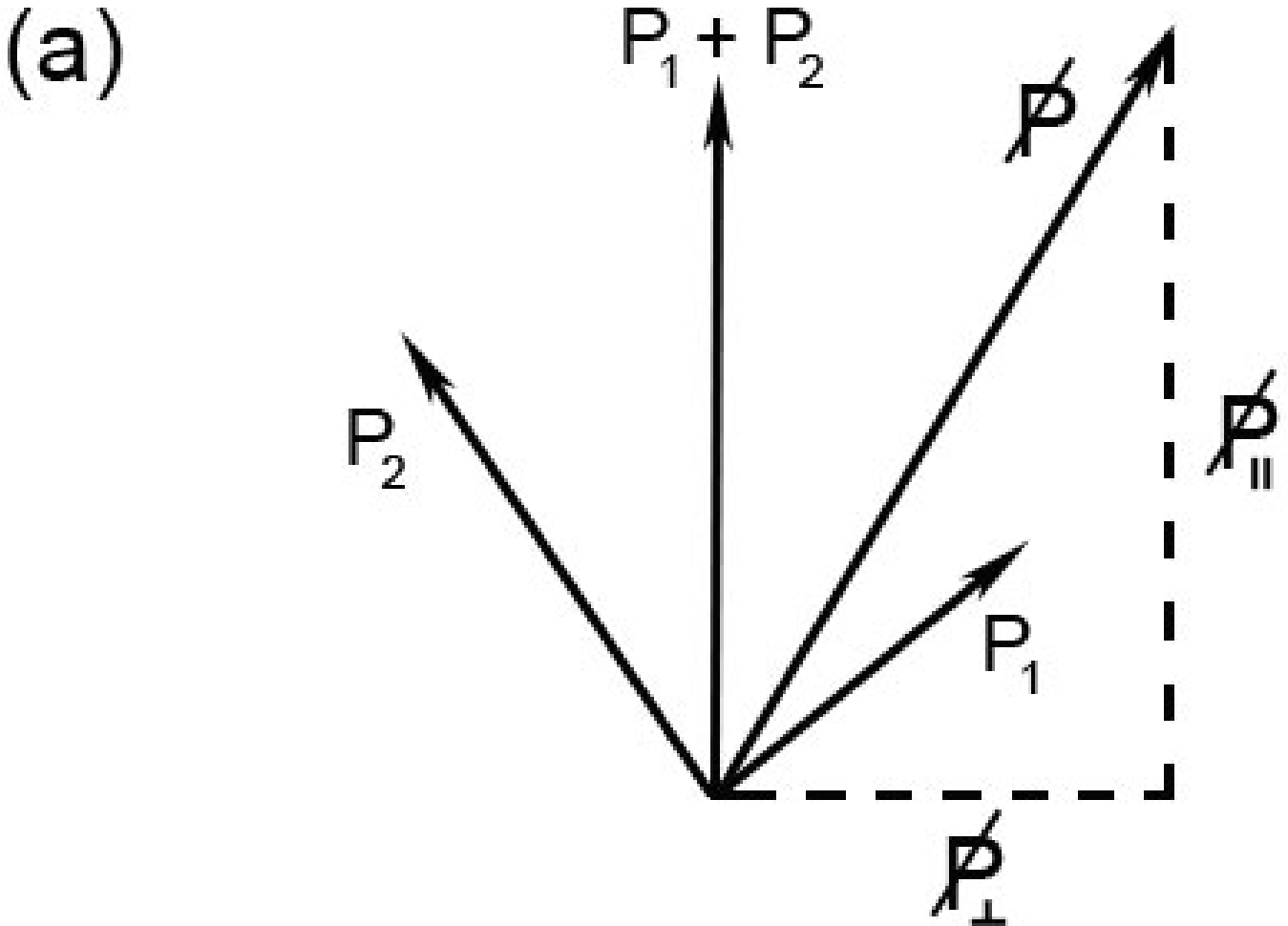}
\includegraphics[width=0.49\linewidth]{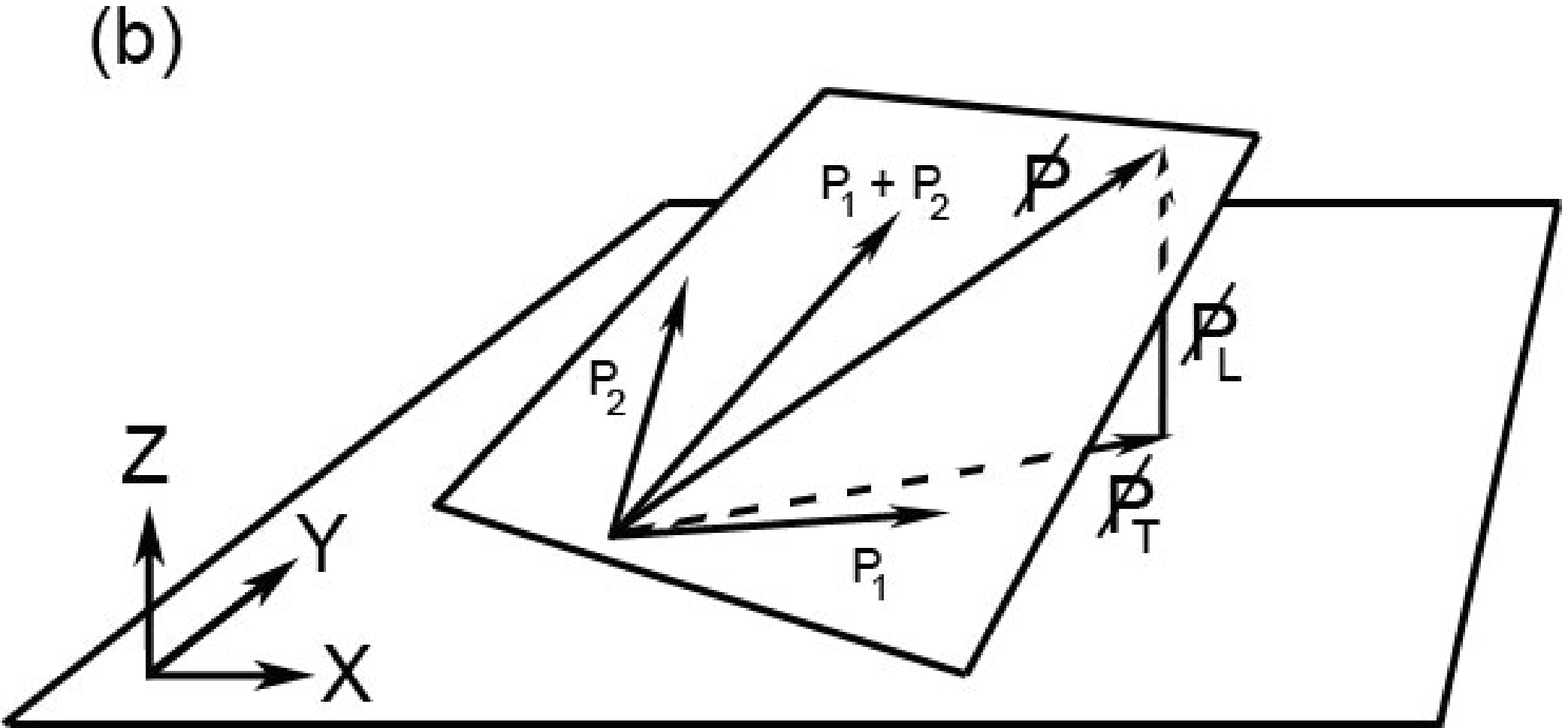}
\end{center}
\caption{\small \emph{ (a) When lepton pairs from the decay chain
$\tilde\chi_i^0\to\ell^\pm\tilde\ell^\mp\to\ell^+\ell^-\tilde\chi_1^0$
have maximal invariant mass, their momenta $P_{1,2}$ in the lab frame
are coplanar with the LSP momentum $\not\!P$. (b) The LSP's
longitudinal momentum $\not\!P_L$ can be found by the condition that
$\not\!P$ is in the plane of the leptons.} }
\label{fig:geom}
\end{figure}

Instead, we will proceed temporarily as if we already knew the
individual $\vec{p}_{\chi,\chi'}^{~T}$ as opposed to just their sum
(\ref{con910b}). Consider, then, just one of the neutralino decays,
say $\tilde\chi_i^0\to e^\pm \tilde e^\mp\to e^+e^-\tilde\chi_1^0$. In
the $\tilde\chi_i^0$ rest frame, the leptons' three-momenta ($\equiv
\vec{P}_{1,2}'$) are back-to-back and collinear with the
LSP's three-momentum ($\equiv \vec{\not\!{P}}'$); thus, in the lab
frame where $\tilde\chi_i^0$ has a velocity $\vec{\beta_1}$, the
boosted momenta $\vec{P}_1$, $\vec{P}_2$, and $\vec{\not\!{P}}$ lie in
the same plane (see Fig.~\ref{fig:geom}a). Looking at this the other
way round, $\vec{\beta_1}$ is the Lorentz boost which makes the
observed lepton momenta antiparallel and fixes the magnitude of their
sum to be $K_i$ --- which we \emph{a priori} don't know at this point
--- {\it i.e.\ } constraints (\ref{con34b}) and (\ref{con7b}). Necessarily,
$\vec{\beta_1}$ must be in the observed leptons' plane. If we choose
a basis in this plane $(\hat{p}_\|,~\hat{p}_\bot)$
parallel/perpendicular\footnote{A potential ambiguity in defining
$\hat{p}_\bot$ is resolved by defining it such that
$\vec{P}_1\cdot\hat{p}_\bot$ is positive:
\begin{equation}
\hat{p}_\bot \equiv \frac{\vec{P}_1 -
(\vec{P}_1\cdot\hat{p}_\|)\hat{p}_\|}{|\vec{P}_1 - (\vec{P}_1\cdot\hat{p}_\|)\hat{p}_\||}.
\nonumber
\end{equation}}
to the total leptonic momentum $\vec{P}\equiv\vec{P}_1 +\vec{P}_2$,
then the boost $\vec{\beta_1} = (\beta_\|,~\beta_\bot)$ must satisfy
three sets of constraints:
\begin{enumerate}
  \item The transformed leptonic momenta must be antiparallel:
  \begin{equation}
  \label{antipar}
  \vec{P}_1' \cdot \vec{P}_2' = - |\vec{P}_1'| |\vec{P}_2'|,
  \end{equation}
  where the transformed four-vectors are given in terms of the boost by
  \begin{equation}
  \label{lepboost}
  \left(
  \begin{array}{c}
    E_{1,2}' \\
    P_{1,2}'^\|  \\
    P_{1,2}'^\bot \\
  \end{array}
  \right)  =
  \left(
  \begin{array}{ccc}
    \gamma & -\beta_\| \gamma & -\beta_\bot \gamma  \\
    -\beta_\| \gamma & 1 + (\gamma-1)\frac{\beta_\|^2}{\beta^2}
    & (\gamma-1)\frac{\beta_\| \beta_\bot}{\beta^2}  \\
    -\beta_\bot \gamma & (\gamma-1)\frac{\beta_\| \beta_\bot}{\beta^2}
    & 1 + (\gamma-1)\frac{\beta_\bot^2}{\beta^2}  \\
  \end{array}
  \right)
  \left(
  \begin{array}{c}
    E_{1,2} \\
    P_{1,2}^\| \\
    P_{1,2}^\bot \\
  \end{array}
  \right),
  \end{equation}
  with $\beta^2\equiv\beta_\|^2+\beta_\bot^2$ and
  $\gamma\equiv (1-\beta^2)^{-1/2}$.
  \item The transformed total leptonic momentum must equal
  $\vec{K}_i=(K_i^\|,~K_i^\bot)$:
  \begin{equation} \label{totlepboost}
  \left(
  \begin{array}{c}
    E' \\
    K_i^\| \\
    K_i^\bot \\
  \end{array}
  \right) =
  \left(
  \begin{array}{ccc}
    \gamma & -\beta_\| \gamma & -\beta_\bot \gamma  \\
    -\beta_\| \gamma & 1 + (\gamma-1)\frac{\beta_\|^2}{\beta^2}
    & (\gamma-1)\frac{\beta_\| \beta_\bot}{\beta^2}  \\
    -\beta_\bot \gamma & (\gamma-1)\frac{\beta_\| \beta_\bot}{\beta^2}
    & 1 + (\gamma-1)\frac{\beta_\bot^2}{\beta^2}  \\
  \end{array}
  \right)
  \left(
  \begin{array}{c}
    E \\
    P \\
    0 \\
  \end{array}
  \right),
  \end{equation}
  where the components $(K_i^\|, K_i^\bot)$ are also unknown at this point.
  \item The inverse-Lorentz-boosted LSP four-momentum must satisfy
  \begin{equation} \label{lspboost}
  \left(
  \begin{array}{c}
    \slashchar{E} \\
    \slashchar{P}_\| \\
    \slashchar{P}_\bot \\
  \end{array}
  \right)  =
  \left(
  \begin{array}{ccc}
    \gamma & \beta_\| \gamma & \beta_\bot \gamma  \\
    \beta_\| \gamma & 1 + (\gamma-1)\frac{\beta_\|^2}{\beta^2}
    & (\gamma-1)\frac{\beta_\| \beta_\bot}{\beta^2}  \\
    \beta_\bot \gamma & (\gamma-1)\frac{\beta_\| \beta_\bot}{\beta^2}
    & 1 + (\gamma-1)\frac{\beta_\bot^2}{\beta^2}  \\
  \end{array}
  \right)
  \left(
  \begin{array}{c}
    \sqrt{K_i^2 + m_1^2} \\
    -K_i^\| \\
    -K_i^\bot \\
  \end{array}
  \right).
  \end{equation}
\end{enumerate}
After some algebra, see the Appendix, these constraints are found to
uniquely determine the unknown boost $(\beta_\|,~\beta_\bot)$ in terms
of the known lab frame leptonic momenta and unknown LSP momenta:
\begin{equation}
\label{betasoln}
\beta_\| = \frac{P}{E(1+ \alpha x)}\quad {\rm and} \quad
\beta_\bot = \alpha \beta_\|,
\end{equation}
where
\begin{equation}\nonumber
\alpha \equiv \frac{\slashchar{P}_\bot}{\slashchar{P}_\|+P}
\quad {\rm and} \quad
x\equiv\frac{P(P_1^\bot-P_2^\bot)(P_2^\|-P_1^\|)}{2E(E_1 P_2^\| + E_2 P_1^\|)}.
\end{equation}
Thus, knowing the leptonic momenta and missing momentum from the LSP
determines $x$ and $\alpha$, hence $\vec{\beta}_1$ and all the
kinematic information in the event. At first glance this may seem
useless since we can only have knowledge of the \emph{transverse}
component of $\vec{\slashchar{P}}$ in the lab coordinate system,
$\vec{\slashchar{P}_T}$, and there are two LSPs that contribute to the
measured total missing momentum.

However, given the transverse component of the LSP momentum we can in
fact reconstruct the longitudinal component $\vec{\slashchar{P}_L}$ by
the following trick: since $\vec{\slashchar{P}}$ must lie in the plane
of the leptons while $\vec{\slashchar{P}_T}$ is by definition in the
transverse $\hat{x}$-$\hat{y}$ plane, $\vec{\slashchar{P}_L}$ must be
of the precise size along $\hat{z}$ to bring $\vec{\slashchar{P}_T} +
\vec{\slashchar{P}_L}$ into the leptons' plane (see
Fig.~\ref{fig:geom}b), {\it
i.e.\ }$(\vec{\slashchar{P}_T}+\vec{\slashchar{P}_L})
\cdot(\vec{P}_1\times\vec{P}_2)=0$, giving
\begin{equation}
\label{plong}
\vec{\slashchar{P}}_L = -\frac{\slashchar{P}_{Tx}(P_{1y}P_{2z} - P_{1z}P_{2y})
+ \slashchar{P}_{Ty}(P_{1z}P_{2x} - P_{1x}P_{2z})}{P_{1x}P_{2y} -
P_{1y}P_{2x}} \hat{z}.
\end{equation}
With both $\vec{\slashchar{P}_T}$ and $\vec{\slashchar{P}_L}$ known we
may immediately project $\vec{\slashchar{P}}$ into the basis
$(\slashchar{P}_\|,~\slashchar{P}_\bot)$,\footnote{If the leptons
happen to be parallel $\vec{\slashchar{P}_L}$ remains undetermined. We
ignore events with this pathological arrangement.}  compute $\alpha$
and insert into (\ref{betasoln}) to solve for the boosts. We can then
use Eq.~(\ref{totlepboost}) to solve for $K_i^\|$ and $K_i^\bot$:
\begin{equation}\label{kparperp}
K_i^\|= P\left(\frac{\gamma+\alpha^2}{1+\alpha^2}-\frac{\gamma}{1+\alpha x}\right),
\quad\quad
K_i^\bot=\alpha P\left(\frac{\gamma-1}{1+\alpha^2}-\frac{\gamma}{1+\alpha x}\right),
\end{equation}
which are also related by $K_i^\bot = \alpha (K_i^\| - P)$. We see
that in the limit $\alpha\to 0$ both $K_i^{\|,\bot}\to 0$, and that
also Eq.~(\ref{betasoln}) correctly reduces to the off-shell result
$\beta=P/E$.

Finally, the LSP mass can then be found from the energy component of
Eq.~(\ref{lspboost}):
\begin{equation}
\label{mlsp}
m_1 = \sqrt{\left(\frac{\slashchar{P}_\|}{\beta_\| \gamma} +
\left(1+ \frac{(\gamma -1) \beta_\|^2}{\beta^2}\right)\frac{K_i^\|}{\beta_\| \gamma}
+\frac{(\gamma - 1) \beta_\bot K_i^\bot}{ \beta^2 \gamma}
\right)^2 - {K_i^\|}^2 - {K_i^\bot}^2},
\end{equation}
which again reduces to the off-shell result of Eq.~(\ref{m1eqn}) when
$K_i^{\|,\bot}\to 0$.  The heavier neutralino mass follows from the
energy component of (\ref{totlepboost}) and energy conservation in its
decay, {\it i.e.}
\begin{equation}\label{m2eqn}
m_i = \sqrt{K_i^2 + m_1^2} + \gamma E - \beta_\| \gamma P,
\end{equation}
while the slepton mass is related to $m_1$ and $m_i$ by
(\ref{con1b}).

Finally, let us return to deal with the realistic situation where we
know only the sum of the LSP momenta
$(\vec{p}_\chi+\vec{p}_{\chi'})^T$. From the discussion above, every
assignment of $\not\!\vec{P}_T=\vec{p}_{\chi}^{~T}$ will yield a set
of masses $\{m_1',~m_i',~m_s'\}$ which satisfy (\ref{con1b}),
(\ref{con34b}), and (\ref{con7b}). Then the other LSP has its
$\vec{p}_{\chi'}^{~T}$ fixed as
$\vec{p}_{\chi'}^{~T}=\not\!\vec{p}^{~T}-\vec{p}_{\chi}^{~T}$,
giving another set of masses $\{m_1'',~m_i'',~m_s''\}$ which satisfy
(\ref{con2b}), (\ref{con56b}), and (\ref{con8b}). Under the
simplifying assumption that the event contains the process in
(\ref{hdecay}) with $i=j$, we should clearly insist that at least
$\{m_1',~m_i'\}\simeq\{m_1'',~m_i''\}$ within some error (we reserve
the possibility that $m_s' \neq m_s''$). This, in principle,
provides two constraints on our choice of the two components of
$\vec{p}_{\chi}^{~T}$, {\it i.e.\ }the system
(\ref{con1b})--(\ref{con910b}) is solved.\footnote{Notice that
something quite interesting has happened here in that we have gotten
around the usual four-fold ambiguity in designating `near' and 'far'
leptons in the decay chains.}

Thus, in the end, we still have to resort to a numerical search for a
solution to (\ref{con1b})--(\ref{con910b}), but this is only over the
two-dimensional space of one of the LSPs' transverse momenta,
$(p_{\chi}^x, ~p_{\chi}^y)$.  Nevertheless, it is not at all obvious
that there won't be multiple solutions with different $\{m_1,~m_i\}$
within a given level of tolerance --- and when one adds to this the
same caveat as in the three-body case of picking events within some
$\epsilon$ of the endpoint, as well as the effects of detector
smearing and issues with backgrounds, it will have to fall to a MC
simulation to test the practicality of the method.

\subsection{Monte Carlo Test}
\label{subsec:mc}

We perform a Monte Carlo study by generating SUSY signal events for
the SPS1a benchmark point~\cite{Allanach:2002nj} using {\tt
PYTHIA~6.413}~\cite{Sjostrand:2006za}, and SM background events with
{\tt HERWIG 6.510}~\cite{Corcella:2000bw,Moretti:2002eu}, interfaced
to {\tt ALPGEN 2.13}~\cite{Mangano:2002ea} for the production of
high jet multiplicities matched to parton showers and {\tt JIMMY
4.31}~\cite{Butterworth:1996zw} for multiple interactions. The
benchmark point is chosen mainly for the sake of comparison with
results obtained with other mass reconstruction techniques, which we
will comment more on in Section \ref{sec:conc}. The generated events
are then put through a fast simulation of a generic LHC detector,
{\tt AcerDET-1.0}~\cite{Richter-Was:2002ch}, widely used to simulate
analyses of high-$p_T$ physics at the LHC. This incorporates such
detector effects as the deposition of energy in calorimeter cells,
and the smearing of electron, photon, muon and hadronic cluster
energies with parameterized resolutions. The {\tt AcerDET-1.0}
isolation requirement for leptons is less than 10 GeV energy in a
$R=0.2$ cone around the lepton and a minimum distance of $\Delta
R=0.4$ from calorimetric clusters. The MC setup is essentially the
same as in~\cite{Kraml:2008zr} and we refer the reader to that paper
for more details. However, we point out that we use $p_T$ dependent
lepton efficiencies based on full simulation studies published
in~\cite{:2008zzm}.

All SUSY processes and relevant SM backgrounds are generated with a
number of events corresponding to an integrated LHC luminosity of
300~fb$^{-1}$. The dominant type of neutralino pairs produced at SPS1a
are $\tilde\chi_2^0 \tilde\chi_2^0$; we will therefore concentrate on
decays of the form $\tilde\chi_2^0(\to e^+ e^-\tilde\chi_1^0)~
\tilde\chi_2^0 (\to \mu^+ \mu^-\tilde\chi_1^0)$, and results in the
previous section can be simplified somewhat by setting $i=j=2$, and in
particular $K \equiv K_i = K_j$.

\begin{figure}[t]
\begin{center}
\includegraphics[width=0.49\linewidth]{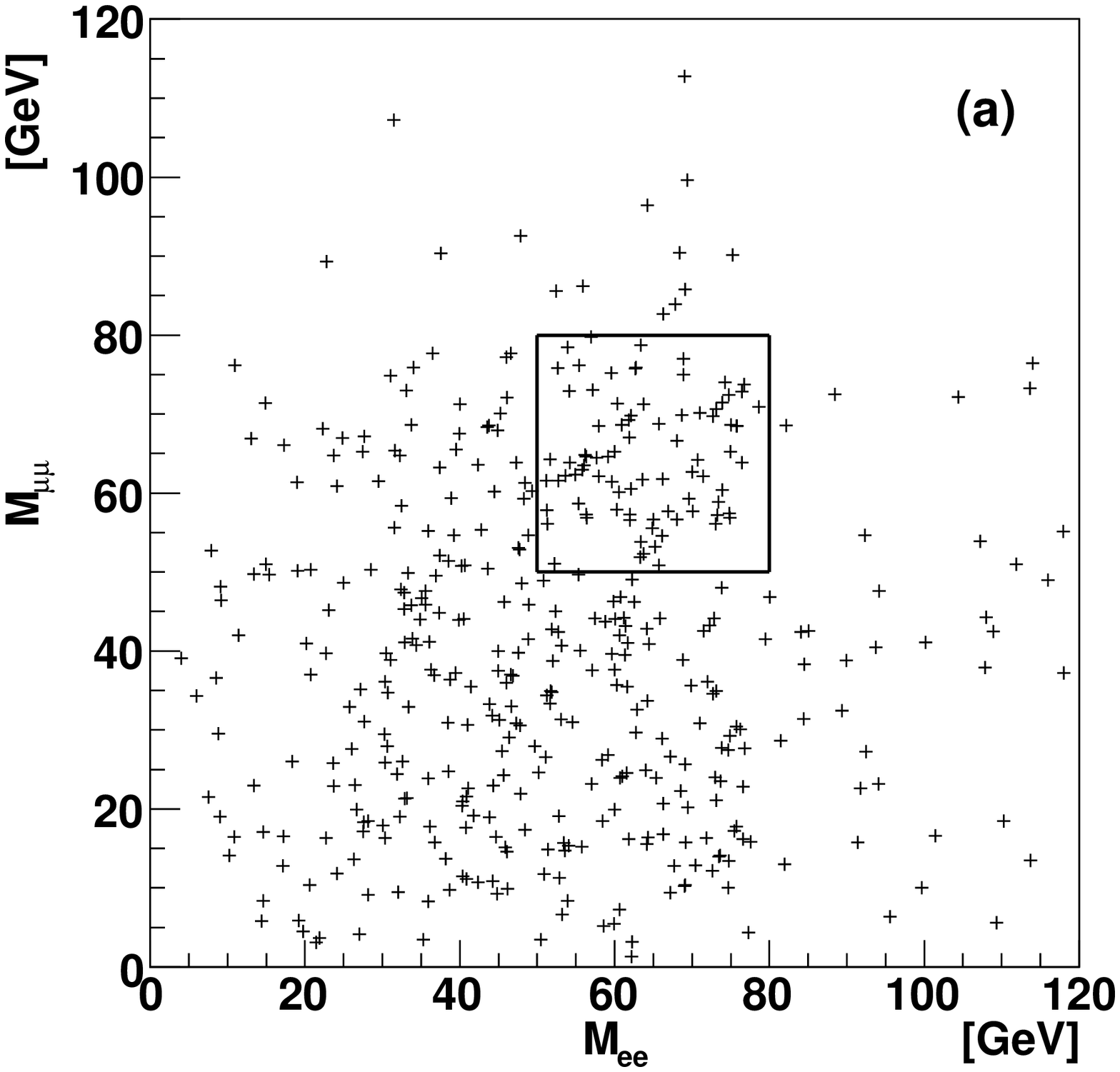}
\includegraphics[width=0.49\linewidth]{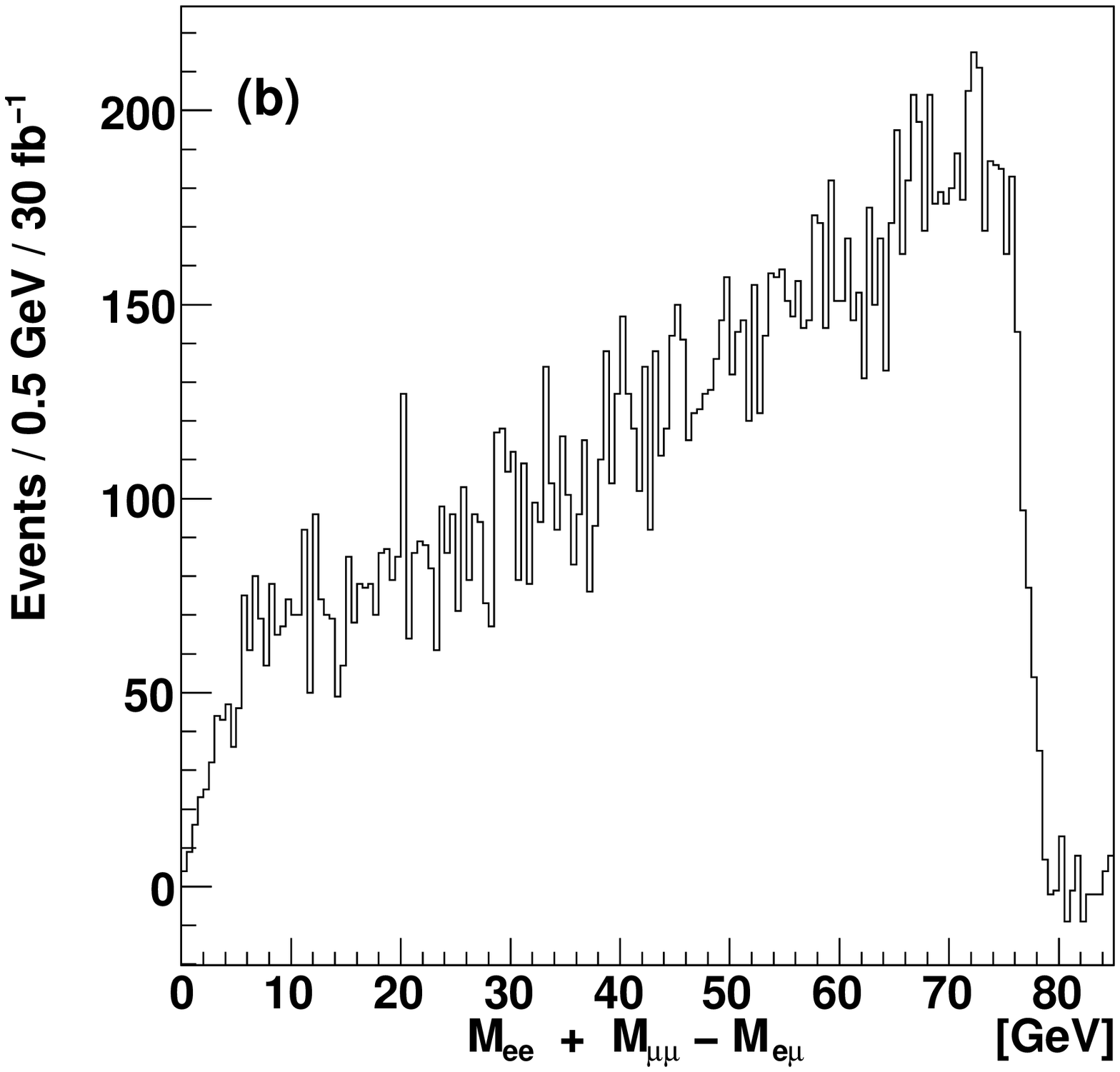}
\end{center}
\caption{\small \emph{ (a) Wedgebox plot for 300~fb$^{-1}$ of integrated
LHC luminosity at the SPS1a benchmark point. Events are sampled from
the boxed region shown for DK analysis. (b) The flavor-subtracted
dilepton invariant mass distribution provides a clean determination of
the edge $M_{\ell\ell}^{max} = 77.1$~GeV. The plot is shown for an integrated luminosity of 30~fb$^{-1}$.} } \label{fig:wb}
\end{figure}

For this analysis we use the same set of cuts as for the NMSSM case in
the previous Section, giving a signal size of roughly 470 events.
Note the requirement of four isolated leptons with flavor structure
$e^+e^-\mu^+\mu^-$ reduces most SM backgrounds to a negligible
level\cite{Bisset:2005rn}. Support for this in the context of a full
simulation of the ATLAS detector is found in Higgs searches for the
channel $h\to ZZ^*\to 4\ell$, {\it e.g.\ }discussed
in~\cite{:1999fr,Aad:2009wy}. The remaining backgrounds of any
importance are found to be $t\bar t$, $Zb\bar b$ and irreducible
$Z^{(*)}Z$. We have simulated a large sample of $t\bar t$ events with
up to two additional hard jets, and find no surviving events with the
additional missing energy and jet cuts. The $Zb\bar b$ and $Z^{(*)}Z$
backgrounds are also expected to be very small after these
cuts. However, because the $Z$ mass is sufficiently far from the
dilepton edge, any remaining events from these backgrounds do not
significantly influence the region of interest in the di-electron
versus di-muon invarant mass plane, shown in the wedgebox plot of
Fig.~\ref{fig:wb}a.\footnote{However, for some SUSY parameter points
one might have a dilepton edge very close to the $Z$ mass.  Though the
$Z$ background events would thus be unavoidably mixed in with signal
events, they would in general not have solutions in the numerical
procedure described in the following. This property of DK gives it a
certain resilience in the face of backgrounds.}

The position of the edge of the box-like structure at
$M_{\ell\ell} \approx 75$~GeV is visually apparent in
Fig.~\ref{fig:wb}a, and, as shown in several studies, can be brought
into precise (sub-GeV) agreement with the nominal value
$M_{\ell\ell}^{\max}=77.07$~GeV, by the standard study of the
flavor-subtracted di-lepton mass distribution shown in
Fig.~\ref{fig:wb}b. One advantage of the DK technique is that we do
not strictly need such precise determination of the edge --- GeV-level
will do to determine our sampling region --- but we will assume that
the edge has been measured to $76.7\pm 0.1$~GeV as quoted
in~\cite{Gjelsten:2004ki}.

Events in a broad neighborhood of the corner of the box,
$M_{ee,\mu\mu}=65\pm \epsilon$~GeV (see further comments below
on sampling regions), are passed to the on-shell DK analysis described
in the previous Section. In detail, the procedure used is the
following:

\begin{enumerate}
  \item An event is selected if the two invariant masses $M_{ee}$
  and $M_{\mu\mu}$ both lie within the $\epsilon$-defined
  region of the wedgebox plot.
  \item A point in $(\slashchar{P}_{Tx},~\slashchar{P}_{Ty})$-space is
  chosen by a uniform scan of $-500$~GeV $< \slashchar{P}_{Tx,y} <$
  $500$~GeV, in $0.2$~GeV steps; the point is assumed to equal the
  transverse momentum of the LSP accompanying the $e^+$ and
  $e^-$ whose four-momenta are $(E,\vec{P})_{1,2}$,
  respectively.
  \item The longitudinal component of the LSP's momentum is found from
  (\ref{plong}).
  \item Components of the LSP momentum in the basis
  parallel/perpendicular to the total leptonic momentum $\vec{P}
  = \vec{P}_1 + \vec{P}_2$ are determined  and used to compute
  $\alpha \equiv \slashchar{P}_\bot /(\slashchar{P}_\| + P)$.
  \item The boost parameters $\beta_\|$ and
  $\beta_\bot$ are now computed from (\ref{betasoln}),
  $K^\|$ and $K^\bot$  from (\ref{kparperp}).
  \item The masses $\{m_1',~m_2'\}$ are computed
  from (\ref{mlsp}) and (\ref{m2eqn}).
  \item Using
  the missing momentum constraint $\vec{p}_{\chi'}^{~T}=
  \not\!\vec{p}^{~T}-\vec{p}_{\chi}^{~T}$, \textbf{steps 3-6} are
  repeated for the LSP accompanying the $\mu^+ \mu^-$ pair,
  obtaining either another set of masses $\{m_1'',~m_2''\}$ or no valid solution for the second mass set.
  \item If no valid second set of masses was obtained, the point is
  assigned zero weight. Otherwise, the two sets of mass solutions are
  plotted with the following weight:
  \begin{equation}\label{likelihood}
  P(\slashchar{P}_{Tx},\slashchar{P}_{Ty})=\frac{1}{\sqrt{2\pi
  \sigma^2}} \exp \left(-\frac{(m_1'-m_1'')^2} {2\sigma^2}
  -\frac{(m_2'-m_2'')^2}{2\sigma^2}\right),
  \end{equation}
  where $\sigma$ is our expectation for the spread between the mass
  values on either side of the event. This should be of the order of
  the missing energy resolution, which is a function of the total
  transverse energy $E_T$ deposited in the calorimeters. We therefore
  assign sigma on an event-by-event basis, using the expected
  performance of the ATLAS detector in SUSY events (see figure 10.84
  of reference ~\cite{:2008zzm}):
  \begin{equation}
  \sigma = 0.57 \sqrt{\sum E_T}.
  \end{equation}
  \item The scan is continued until all points have been assigned a weight.
  \item The procedure is repeated for all events that have invariant masses sufficiently close to the endpoint.
  \end{enumerate}

Our final mass distribution for a single event is obtained by
histogramming all mass solutions found in the scan over missing
momentum components, weighted by Eq.~(\ref{likelihood}). For events
that lie \emph{exactly} at the endpoint, and where the mismeasurement
of lepton momenta and missing energy is negligible, the peak of the
two mass distributions for $\{m_1',~m_2'\}$ and $\{m_1'',~m_2''\}$
should coincide, agreeing with the nominal values of the masses. We
find that this remains accurate for events that lie sufficiently close
to the endpoint, and thus with a value of $\epsilon$ that is not too
large.

In fact, choosing a value of $\epsilon$ is essentially a trade-off
between accumulating statistics by allowing more events to pass the
cut, thus reducing the fluctuations that come from the smearing of
lepton and missing energy momenta, and protecting the integrity of the
mass solutions that are obtained by the procedure at $\epsilon=0$. We
find that a compromise value for $\epsilon$ of 15 GeV gives just
enough events with peak region close to the nominal values to dominate over events that display
either one or two degenerate peaks or peaks in the wrong place. In the latter case, we observe that
the maximum weight at the peak is lower.

\begin{figure}[t]
\begin{center}
\includegraphics[width=0.7\linewidth]{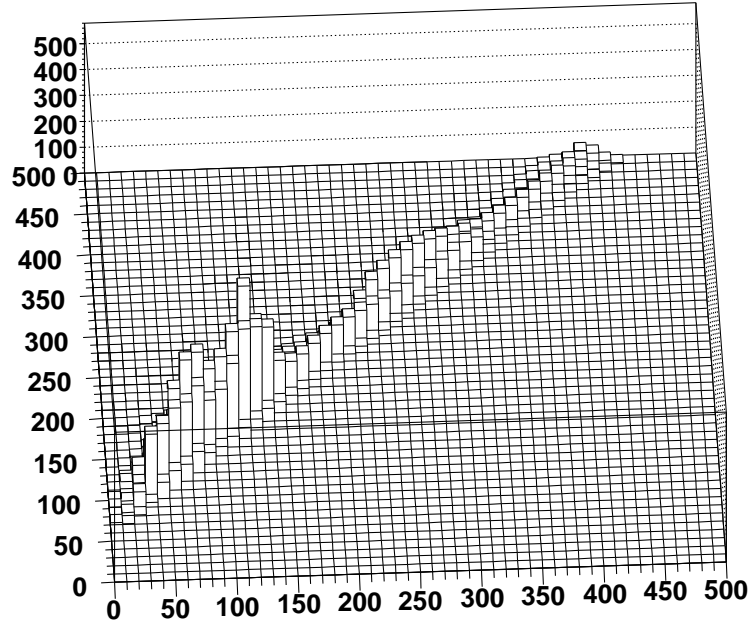}
\end{center}
\caption{\small \emph{Mass distribution for 300 fb$^{-1}$ of data at SPS1a,
obtained using the procedure described in the text. The contribution
of individual events with degenerate solutions is still clearly
visible, although it is a region close to the nominal mass values of
$(m_1,m_2)=(96.1,176.8)$~GeV that emerges with the largest total
weight.}}
\label{fig:finalmasses}
\end{figure}

By summing the distributions for events within our choice of
$\epsilon$, the resulting total mass distribution for 300 fb$^{-1}$ of
data is shown in Fig.~\ref{fig:finalmasses}. Although the existence of
multiple peaks is clear with the limited statistics, the point with
the largest total weight is found to be
$(m_1,m_2)=(114.75,191.75)$~GeV, close to the nominal value of
$(m_1,m_2)=(96.1,176.8)$~GeV. We emphasize that this procedure is
simply a suggestion for an estimator of the masses, and although there
are similarities in shape, Eq.~(\ref{likelihood}) is not a likelihood
function.

To check the robustness of the estimator and find the statistical
error on making such a measurement, we have performed 10 independent
`experiments' with 300 fb$^{-1}$ integrated luminosity. In each
case, the mass solution with the largest total weight fell near the
nominal masses, with a standard deviation of 20.2 GeV on $m_1$ and
21.2 GeV on $m_2$. While these errors are quite large, there is
undoubtedly scope for improvement. By better understanding the
properties of events with degenerate or wrong solutions one could
search for a system of kinematic cuts to remove these subsets; one
could also increase statistics to tighten the cut on $\epsilon$ or
investigate other estimators for the masses with better properties
with respect to these events.

Incidentally, had we wrongly assumed off-shell kinematics at this
parameter point, and hence tried using the technique of Section
\ref{sec:off} to analyze events in the boxed region of
Fig.~\ref{fig:wb}a, we would have failed since essentially no events
provide two mass solutions with a near equal mass.  This provides a
way of distinguishing a sample of on- versus off-shell
decays\footnote{One exception occurs if it happens that $K=0$,
{\it i.e.\ } $m_s = \sqrt{m_1 m_2}$. Then both on-shell and off-shell
techniques are equally applicable and should reconstruct the same LSP
mass. At SPS1a, $K\approx 18$~GeV, which is fairly small compared to
the maximum $K$ one can get with the same neutralino masses, $K^{\max}
= \frac{m_2}{2}(1 - m_1^2/m_2^2)\approx 62$~GeV, but safely away from
zero.} distinct from the usual way of measuring the departure of the
di-lepton mass distribution from a triangular
shape~\cite{Kraml:2008zr}, or looking for specific relationships
between the positions of lepton-jet invariant mass
maxima~\cite{Lester:2006cf}.

Finally, note that this method, although we have not explicitly
demonstrated it, is in principle also applicable to extracting the
slepton mass.

\section{Conclusions}
\label{sec:conc}

This paper completes the demonstration of the application of the DK
technique to neutralino decays in SUSY models, or indeed any similar
decay chain in models such as {\it e.g.\
}UED~\cite{Appelquist:2000nn}, having explicitly shown the procedure
for reconstructing both off- and on-shell decays to lepton pairs in
realistic MC simulations of two very different supersymmetry
benchmark points. We find that in the three-body decay scenario we
can reconstruct the LSP mass with an accuracy of around 4 GeV, while
the decay through an on-shell slepton allows a precision of 20 GeV.

DK has the advantage of simplicity and robustness in the face of
backgrounds; for non-signal events there tends to be no solution for
the constraint equations, or at least no \emph{preferred} solution
in the distribution of events passing loose constraint requirements.
Moreover, since the method only makes use of unlike di-leptons
($e^+e^-\mu^+\mu^-$) from the $\tilde\chi^0_2\to\tilde\chi^0_1$
transitions, it is insensitive to the combinatoric issues that arise
when one considers particles produced further up the decay chain. It
may hence complement mass determinations which exploit the full
chains such as \cite{Cheng:2009fw,Webber:2009vm}. Generalization to
other decaying states ({\it e.g.} charginos as
in~\cite{Kersting:2008qn}), perhaps using jets instead of or in
addition to leptons, is open to investigation.

Perhaps the major disadvantage of DK is the requirement of a high
event rate: one typically needs at least ${\mathcal O}(10)$ events in
the neighborhood of a kinematic endpoint to make the reconstruction
stable, and, in the case of four-lepton final states, that translates
to several hundreds of events needed on a wedgebox plot. In addition,
the case of on-shell decays gives rise to extra solutions in the mass
space that can be eliminated with more statistics, but ultimately
contribute to an increased error in the reconstructed masses.

Thus, in the case of neutralino-pair production considered in this
work, DK may, depending on the parameter point Nature has chosen,
perhaps only serve useful as a check on results obtained with other
techniques that do not depend on events near an endpoint. As
mentioned in the Introduction, these most prominently include MSTs
and $m_{T2}$ techniques, which at the SPS1a point happen to work
quite well.

\section*{Acknowledgements}
This work was funded in part by the Kavli Institute for Theoretical
Physics (Beijing). ARR and MJW acknowledge funding from the UK Science
and Technology Facilities Council (STFC). This work is also part of
the French ANR project ToolsDMColl, BLAN07-2-194882.

\newpage

\section*{Appendix}
\subsection*{On-Shell Kinematics}
Consider the neutralino decay
\begin{equation}
\tilde\chi_i^0\to \ell^+\tilde{\ell}^-\to \ell^+\ell^- \tilde\chi_1^0,
\end{equation}
in the neutralino rest frame. If the di-lepton invariant mass is
maximal, all the decay products must be collinear (say along
$\hat{x}$). In particular, four-momentum conservation forces
\begin{equation*}
    \left(
  \begin{array}{c}
    E_{\ell^+} \\
    \\
    p_{\ell^+}   \\
  \end{array}
\right)_{\tilde\chi_i^0} = \left( \begin{array}{c}
    \frac{m_i^2 - m_s^2}{2 m_i} \\
    \\
    \frac{m_i^2 - m_s^2}{2 m_i}\\
  \end{array}
\right)
\quad{\rm and}\quad
  \left(
  \begin{array}{c}
    E_{\tilde{\ell}^-} \\
    \\
    p_{\tilde{\ell}^-} \\
  \end{array}
\right)_{\tilde\chi_i^0} = \left( \begin{array}{c}
    \frac{m_i^2 + m_s^2}{2 m_i} \\
    \\
    -\frac{m_i^2 - m_s^2}{2 m_i}\\
  \end{array}
\right),
\end{equation*}
where $m_s$ is the slepton mass and the lepton is assumed to be
massless. Similarly, in the slepton's decay frame,
\begin{equation*}
    \left(
  \begin{array}{c}
    E_{\ell^-} \\
    \\
    p_{\ell^-}   \\
  \end{array}
\right)_{\tilde{\ell}} = \left( \begin{array}{c}
    \frac{m_s^2 - m_1^2}{2 m_s} \\
    \\
    -\frac{m_s^2 - m_1^2}{2 m_s}\\
  \end{array}
\right)
\quad{\rm and}\quad
  \left(
  \begin{array}{c}
    E_{\tilde\chi_1^0} \\
    \\
    p_{\tilde\chi_1^0} \\
  \end{array}
\right)_{\tilde{\ell}} = \left( \begin{array}{c}
    \frac{m_s^2 + m_1^2}{2 m_s} \\
    \\
    \frac{m_s^2 - m_1^2}{2 m_s}\\
  \end{array}
\right),
\end{equation*}
which, when boosted back to the $\tilde\chi_i^0$ rest frame using
$\beta=-(m_i^2 - m_s^2)/(m_i^2 + m_s^2)$, becomes
\begin{equation*}
    \left(
  \begin{array}{c}
    E_{\ell^-} \\
    \\
    p_{\ell^-}   \\
  \end{array}
\right)_{\tilde\chi_i^0} = \left( \begin{array}{c}
    m_i\frac{m_s^2 - m_1^2}{2 m_s^2} \\
    \\
    -m_i\frac{m_s^2 - m_1^2}{2 m_s^2}\\
  \end{array}
\right)
\quad{\rm and}\quad
  \left(
  \begin{array}{c}
    E_{\tilde\chi_1^0} \\
    \\
    p_{\tilde\chi_1^0} \\
  \end{array}
\right)_{\tilde\chi_i^0} = \left( \begin{array}{c}
    \frac{m_s^4 + m_i^2 m_1^2}{2 m_i m_s^2} \\
    \\
    \frac{m_s^4 - m_i^2 m_1^2}{2 m_i m_s^2}\\
  \end{array}
\right).
\end{equation*}
From these equations it is easy to verify (\ref{con1b}), (\ref{con2b}),
(\ref{con7b}), and (\ref{con8b}).

\subsection*{Finding the Lorentz Boost Parameters}

Here we indicate in more detail how one may arrive at
Eq.~(\ref{betasoln}). Starting from Eq.~(\ref{lspboost}) we have two
relevant equations,
\begin{eqnarray*}
\slashchar{P}_\|  + K_i^\| &=&  \beta_\| \left( \gamma\sqrt{K_i^2 + m_1^2}- (\gamma-1) \frac{\beta_\|}{\beta^2} K_i^\|  -
 (\gamma-1) \frac{\beta_\bot}{\beta^2} K_i^\bot \right),  \\ \label{meq2}
\slashchar{P}_\bot + K_i^\bot  &=&  \beta_\bot \left( \gamma\sqrt{K_i^2 + m_1^2}- (\gamma-1) \frac{\beta_\|}{\beta^2} K_i ^\|  -
 (\gamma-1) \frac{\beta_\bot}{\beta^2} K_i^\bot \right).
\end{eqnarray*}
Taking the ratio of these one obtains after some rearranging
\begin{equation}
\label{keqn}
K_i ^\| = \frac{\beta_\|}{\beta_\bot}\slashchar{P}_\bot + \frac{\beta_\|}{\beta_\bot} K_i^\bot -\slashchar{P}_\|.
\end{equation}
This can now be inserted into two of the equations of
(\ref{totlepboost}),
\begin{eqnarray}  \label{eq1}
K_i ^\|  &=&  -\beta_\| \gamma E+ P+ P(\gamma-1)\frac{\beta_\|^2}{\beta^2} ,\\ \label{eq2}
K_i^\bot &=&  -\beta_\bot \gamma E + (\gamma-1)\frac{\beta_\| \beta_\bot}{\beta^2}P,
\end{eqnarray}
which can then be used to solve for the ratio
\begin{equation}\label{betaratio}
\alpha \equiv \frac{\beta_\bot}{\beta_\|}
= \frac{\slashchar{P}_\bot}{P + \slashchar{P}_\|}.
\end{equation}
Starting from the antiparallel condition (\ref{antipar}) and
expanding with the lepton momenta from (\ref{lepboost}), one
arrives, after some algebra, at
\begin{eqnarray*}
0 = (1+ \beta_\|^2 + \beta_\bot^2)E_1 E_2 - 2 \beta_\|E_1 P_2^\| -2 \beta_\bot E_1 P_2^\bot - 2 \beta_\| E_2 P_1^\|
-2 \beta_\bot E_2 P_1^\bot + \\ 2 \beta_\| \beta_\bot (P_1^\| P_2^\bot +  P_2^\| P_1^\bot) + (1+ \beta_\|^2 - \beta_\bot^2)P_1^\| P_2^\|
+ (1 - \beta_\|^2 + \beta_\bot^2) P_1^\bot P_2^\bot,
\end{eqnarray*}
using that $E_{1,2} = |\vec{P}_{1,2}|$. Substituting $\beta_\bot$ for
$\alpha$ and $\beta_\|$ from (\ref{betaratio}), we get a quadratic
equation for $\beta_\|$:
\begin{equation*}
0 = a \beta_\|^2 - 2b \beta_\| + c,
\end{equation*}
where
\begin{eqnarray*}
a & \equiv & E_1 E_2 + P_1^\| P_2^\|- P_1^\bot P_2^\bot
+ 2\alpha (P_1^\| P_2^\bot +  P_2^\| P_1^\bot)
+ \alpha^2 (E_1 E_2 - P_1^\| P_2^\| + P_1^\bot P_2^\bot), \\
b & \equiv & E_1 P_2^\| + E_2 P_1^\| + \alpha(E_1 P_2^\bot + E_2 P_1^\bot),\\
c & \equiv &  E_1 E_2 + P_1^\| P_2^\| + P_1^\bot P_2^\bot.
\end{eqnarray*}
Using again that $E_{1,2}^2 ={P_{1,2}^\|}^2+{P_{1,2}^\bot}^2$, one
can show that $b^2 -4 a c = 0$, so this quadratic has a double root
\begin{equation}
\nonumber
\beta_\| = \frac{E_1P_2^\|+E_2 P_1^\| + \alpha (E_1 P_2^\bot + E_2 P_1^\bot)}
{E_1E_2+P_1^\|P_2^\|-P_1^\bot P_2^\bot+2\alpha(P_1^\|P_2^\bot+P_1^\bot P_2^\|)
+\alpha^2 (E_1 E_2 - P_1^\| P_2^\| + P_1^\bot P_2^\bot)}.
\end{equation}
Making repeated use of that, by definition, $P_1^\bot = - P_2^\bot$
and $P =P_1^\| + P_2^\|$, one arrives at the equation for $\beta_\|$
in (\ref{betasoln}).

\bibliographystyle{h-physrev4}
\bibliography{onshell}

\end{document}